\pdfoutput=1
\documentclass[runningheads]{llncs}
\usepackage{graphicx}
\usepackage{algorithm}
\usepackage{algorithmic}
\usepackage{multirow}
\begin{document}
\title{Parallelizing Convergent Cross Mapping Using Apache Spark}
% \author{Bo Pu\inst{1} \and
% Lujie Duan\inst{1} \and
% Nathaniel D. Osgood\inst{1}}
\author{Bo Pu \and
Lujie Duan \and
Nathaniel D. Osgood}
\authorrunning{B. Pu et al.}
\institute{Department of Computer Science, Univ. of Saskatchewan, Saskatoon, SK, Canada \\
\email{\{bo.pu, lujie.duan, nathaniel.osgood\}@usask.ca}}
\maketitle              % typeset the header of the contribution
\begin{abstract}

Identifying the causal relationships between subjects or variables remains an important problem across various scientific fields. This is particularly important but challenging in complex systems, such as those involving human behavior, sociotechnical contexts, and natural ecosystems.  By exploiting state space reconstruction via lagged embedding of time series, convergent cross mapping (CCM) serves as an important method for addressing this problem. While powerful, CCM is computationally costly; moreover, CCM results are highly sensitive to several parameter values.  While best practice entails exploring a range of parameter settings when assessing casual relationships, the resulting computational burden can raise barriers to practical use, especially for long time series exhibiting weak causal linkages.  We demonstrate here several means of accelerating CCM by harnessing the distributed Apache Spark platform.  We characterize and report on results of several experiments with parallelized solutions that demonstrate high scalability and a capacity for over an order of magnitude performance improvement for the baseline configuration. Such economies in computation time can speed learning and robust identification of causal drivers in complex systems.
\end{abstract}
\section{Introduction}
Identification of causal relations between variables in many domains has traditionally relied upon controlled experimentation, or investigation of underlying mechanisms.  The first of these requires a heavy investment of  time and financial resources, and can pose ethical concerns. The limits of such controlled studies -- such as Randomized Controlled Trials (RCTs) -- are particularly notable in the context of complex systems, which exhibit reciprocal feedbacks, delays and non-linearities \cite{luke2012systems}.  While ubiquitous in science, progress in the latter approach is commonly measured in decades.  In recent years, a growing number of researchers have applied Pearl's causal inference framework \cite{pearl2009causal} to identify causal linkages, but such methods can be challenging to apply in the presence of observability constraints, and when the variables of interest are coupled but distant within the system. Most critically, in the context of complex systems, such inference techniques encounter challenges in the context of reciprocal causality.

Convergent cross mapping \cite{sugihara2012detecting} is an algorithm based on Takens' embedding theorem \cite{takens1981detecting} that can detect and help quantify the relative strength of unidirectional and bidirectional causal relationships between variables X and Y in coupled complex systems. Within the CCM algorithm, in order to assess if variable $Y$ is causally governed by variable $X$, we attempt to predict the value of $X$ on the basis of the state space reconstructed from $Y$ (see below); for statistical reliability, this must be done over a large number $r$ of realizations.  To assess causality, we examine whether these results "converge" as we consider a growing numbers $L$ of datapoints within $Y$ within our reconstruction (also see below).

Overall, the results of CCM are sensitive to the parameters below:

- $r$ The number of random subsamples, commonly 250 or larger.

- $\tau$ The embedding delay used in the shadow manifold reconstruction. 

- $E$ The embedding dimension of the dynamic system. For simplex projection, $E$ will range from 1 to 10. 

- $L$ The size of the library extracted from time series. 

Running CCM with a wide range of different parameter settings imposes a high computational overhead. But, as for many data science tasks, we believed that the performance could be elevated via parallel and distributed processing by implementing the CCM algorithm atop Apache Spark \cite{zaharia2010spark} (henceforth, ``Spark") and distributing computations across a Yarn cluster \cite{vavilapalli2013apache}. 

In this paper, following additional background on CCM and the related literature, we describe a CCM parallel implementation which utilizes the MapReduce framework \cite{dean2008mapreduce} provided by Spark.  The paper then presents a performance evaluation and comparison of the framework. We can conclude from the experiments that, with the parallel techniques and cloud computing support, researchers can use CCM to confidently infer causal connections between larger time series in far less time than previously required.

\section{Background}

\subsection{Convergent Cross Mapping Basics}
% propose
In 2012, Sugihara et al. \cite{sugihara2012detecting} built on ideas from Takens' Theorem \cite{takens1981detecting} to propose convergent cross mapping (CCM) to test causal linkages between nonlinear time series observations. This approach has enjoyed varied applications. For example, Luo et al. \cite{luo2014causal} successfully revealed underlying causal structure in social media and Verma et al. \cite{verma2016analysis} studied cardiovascular and postural systems by taking advantages of this algorithm. 

We provide here a brief intuition for why and how CCM works.  Consider two variables $X$ and $Y$, each associated with eponymous time-series and -- further -- where $Y$ depends on $X$.
%mutually interacting in a dynamic system
For example, consider a case where for each timepoint $X$ measures the count of hares, and $Y$ that of lynx. In this situation, 
%the dynamics of $X$ (e.g., a rapid rise or persistent drop in the hare population) will often tell us much about the state of other areas of the system that govern it, including  $Y$ (e.g., that there are likely to be few or many lynx around, respectively). 
if $Y$ (lynx) causally depends on $X$ (hares), observing the values of $Y$ over time (e.g., a steep drop or a plateauing in lynx numbers) tells us about the state of governing factors, including $X$ (here, the fact that the number of hares is too small to effectively feed the lynx population, or that they are roughly in balance with lynx, respectively). 
A implication of this -- captured by Takens' Theorem -- 
%is that a space reconstruction based on $Y$ alone should capture information on $X$. For example, if $Y$ is indeed governed by $X$, 
is that information on the state of $X$ is encoded in the state space reconstructed from $Y$, meaning that points that are located nearby within  $Y$'s reconstructed state space
will be associated with similar values for $X$, and can thus be used to make accurate (skillful) prediction of the value of $X$.  In most cases, such prediction of one variable (e.g., $X$) within the state space of another ($Y$) can be achieved by nearest neighbor forecasting using simplex projection \cite{heylen2011fully}. Pearson correlation between observed and predicted values can be applied to measure prediction skill. 

%In order to predict whether $Y$ is causally governed by $X$, the reconstructed states based on $Y$ can be used to cross predict $X$ -- i

% A key need within CCM is to distinguish \textit{causal} dependence of one variable on another from purely statistical dependence (e.g., due to covariation) -- which might also support high prediction skill.  The two can be distinguished by assessing how prediction skill changes as the number data points used in the embedding -- the so-called \textit{library size} $L$ -- rises.  To assess whether $Y$ is causally governed by $X$, we evaluate how successively larger counts $L$ of datapoints in $Y$ change the skill with which values of $X$ can be predicted. If such prediction skill rises monotonically with $L$ it indicates that $X$ is causally driving $Y$ (putting aside certain exceptional cases).  By contrast, the presence of a merely statistical dependence of $Y$ on $X$ may lead to a high basal level of prediction skill, but will not lead to such convergence (such monotonic rise in prediction skill) as $L$ rises.  In order to confidently assess this convergence with rising $L$ in light of stochastic selection of the library, the prediction skill must be assessed over $r$ random subsamples of the time series for each value of $L$. The speed of convergence and the magnitude of the achieved prediction skill for large $L$ indicates the existence and strength of the causal dependence of $Y$ on $X$ in the context of the assumed values for $E$ and $\tau$. 

\subsection{Past work in CCM Performance Improvement}
Despite the fact that CCM is increasingly widely applied, there remain pronounced computational challenges in applying the tool for moderate and large time series. In order to secure confidence in inferences regarding causality, use of appropriate parameter values and a relative longer input are required for the original CCM \cite{monster2017causal}. As such, since its first appearance in 2012, a number of modifications and improvements have been proposed to handle this drawback. In 2014, Ma et al. \cite{ma2014detecting} developed cross-map smoothness (CMS) based on CCM which has the advantage of requiring a shorter time series. Compared to original CCM, CMS can be used for time series in the order of $n = 10$, whereas CCM requires time-series at least in the order of $n = 10^3$ to yield reliable results. Additionally, works such as \cite{cao1997practical}, \cite{kantz2004nonlinear}, \cite{kugiumtzis1996state} investigated and introduced mathematical methods to properly estimate parameters required by CCM (embedding dimension $E$, time delay $\tau$ and time subsample $L$). Such work expanded CCM-related research and also provided methods for quickly inferring causality in certain circumstances.

The previous improvements on CCM typically trade off potential accuracy for relatively fast execution, and the assumptions in some methods cannot be safely maintained with noisy time series observations. However, the original CCM can be improved by introduction of parallel computing techniques. In recent years, numerous studies such as \cite{maillo2017knn}, \cite{reyes2015big} have been conducted using distributed computing frameworks such as MPI or Spark. Such parallel techniques can dramatically improve the algorithmic performance by effectively exploiting the cluster-based computational capacity. It is worthwhile to implement a parallel version of CCM to allow researchers to rapidly and robustly evaluate the existence and strength of causal connections between measured time series.

\section{Methodology}
\label{section:methods}

% introduce data structure here: DataFrame: named columns
% introduce pipeline here
To achieve a Spark parallel version of CCM, we introduce two core concepts: the Spark Resilient Distributed Dataset (RDD) \cite{zaharia2012resilient} and Pipeline. The former is the immutable data structure that can be operated in a distributed manner, which brings significant benefits for concurrently draw $r$ subsamples of time series to assess Cross-Mapping convergence. As for the pipeline, it is specified as a sequence of stages, and each stage transforms the original RDD to another RDD accordingly. In summary, the definition of pipeline supports an elegant design for a parallel CCM algorithm manipulating RDDs in Spark.

\subsection{CCM Transform Pipeline}

\begin{figure}[h]
\centering
\includegraphics[height=3cm]{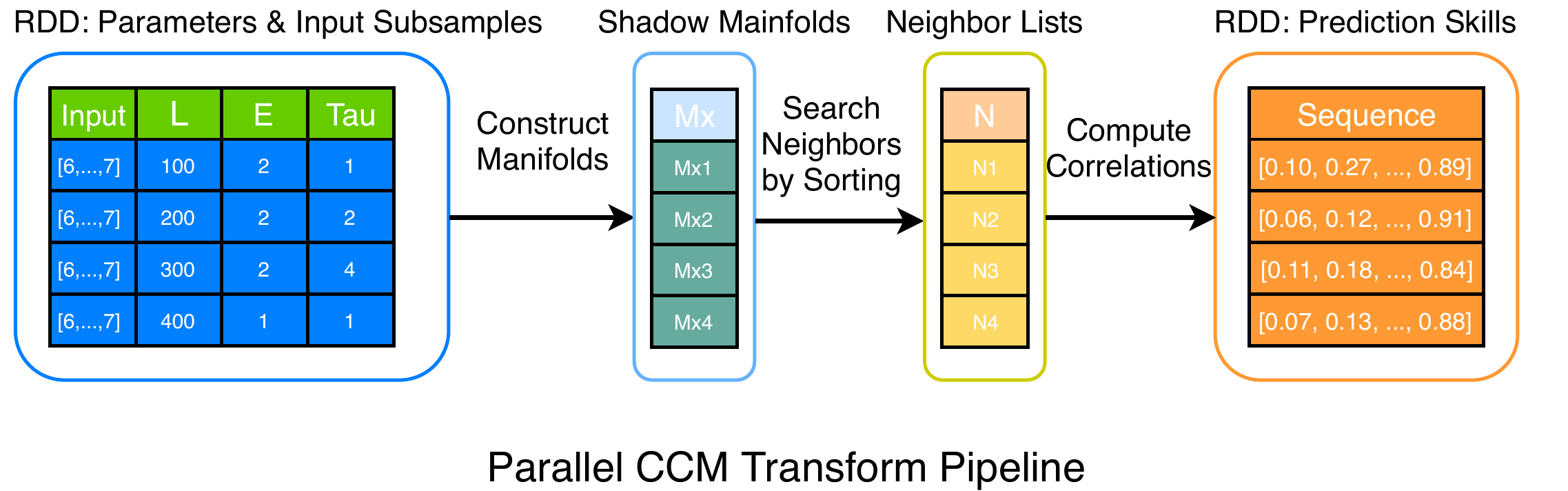}
\caption{A diagram of CCM RDD transformation which takes multiple realizations as input and outputs prediction skills.}
\label{fig:transformation}
\end{figure}

Consider applying CCM to test if the variable associated with time series $Y$ is being driven by the variable associated with time series $X$.  In the corresponding transform pipeline, the parallel version of CCM is implemented as several stages to transform the RDD of $r$ random subsamples of the time series to the RDD of prediction skills for a given ($\tau$, $E$, $L$) tuple. To start the transformation, an input RDD is created which includes a pair of subsamples of lengths $L$ of each of the time series, and values for each of two parameters ($\tau$, $E$). The output of the CCM transform pipeline is an RDD of sequences of prediction skills. In the whole procedure, Spark operates the whole transformation in parallel without extra coding as shown in Fig.~\ref{fig:transformation}.

\subsection{Distance Indexing Table Pipeline}

The CCM transform pipeline above achieves the aim of running CCM concurrently on multiple subsamples $r$. However, there is still a considerable potential for further optimization for this pipeline. As mentioned before, the most time-consuming part in the original CCM is finding the $E+1$ nearest neighbors for every lagged-coordinate vector ($\tau$) in the shadow manifold. For every point in the input RDD, the CCM transform pipeline computes the distances to all lagged-coordinate vectors of subsamples, sorts them and finally takes the top $E+1$ as the nearest neighbors. This process is inefficient because of its repeated sorting and calculation for all the subsamples. It is particularly notable that as the length of subsamples $L$ used for computation increases, the running time will grow superlinearly.

\begin{figure}
\centering
\includegraphics[height=5cm]{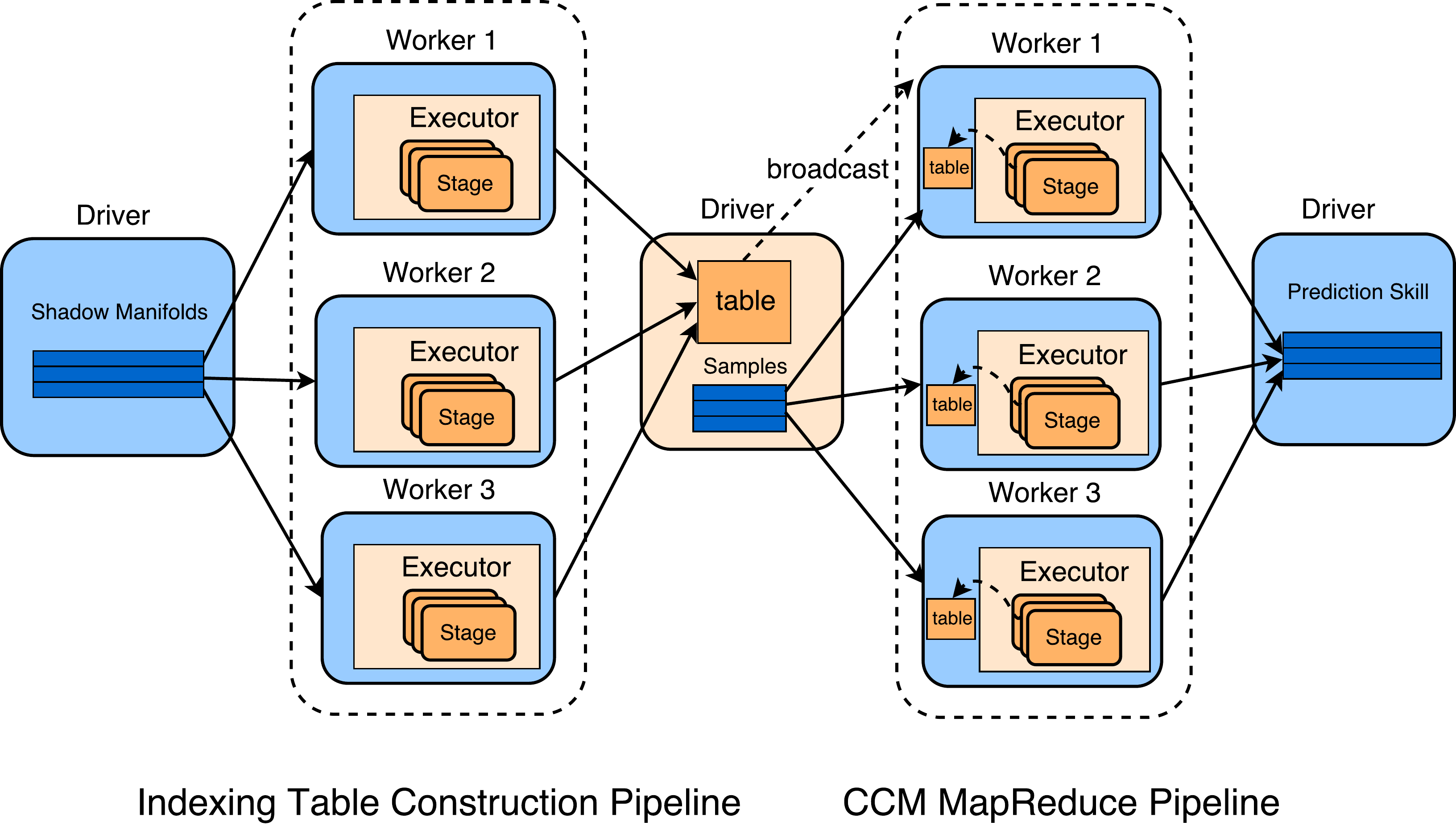}
\caption{An illustration of the dependencies of two pipelines. After the distance indexing table is constructed in parallel, Spark will broadcast it to all nodes and in the next pipeline, the executors can look up the table and fetch $E+1$ nearest neighbors quickly.}
\label{fig:twopipelines}
\end{figure}

The best approach is to break down the nearest neighbors searching of CCM transformations into two parts: distance indexing table construction and nearest neighbors searching based on the constructed table. The first part can be achieved by setting another pipeline as a preprocessing step before applying the CCM transform pipeline. After building the distance indexing table, Spark can broadcast this table to each worker node on the cluster at one time rather than ship a copy of it every time they need it, as shown in Fig.~\ref{fig:twopipelines}. The pipeline of constructing the distance indexing table will be executed concurrently on the entire input time series, and it also reduces a significant amount of repeated calculation in the CCM transform pipeline. From the experiment results, the total computation time decreases in a pronounced fashion. As the library size $L$ grows, the time spent on searching for the nearest neighbors increases correspondingly, and pre-building the distance indexing table secures increasing benefit.

% As mentioned in the above, examining prediction skill for differing values of $L$ assists to qualify prediction skills convergence. Thus, experimenting with a wide range of $L$ to assess the causality is desirable. Considering that two other parameters ($E$ and $\tau$) are typically small values (commonly less than or equal to 10) used in simplex projection, speeding up this algorithm for large values of $L$ situation is of most important.

\subsection{Asynchronous Pipelines}

% handling asynchronous computation
\begin{figure}
\centering
\includegraphics[height=3cm]{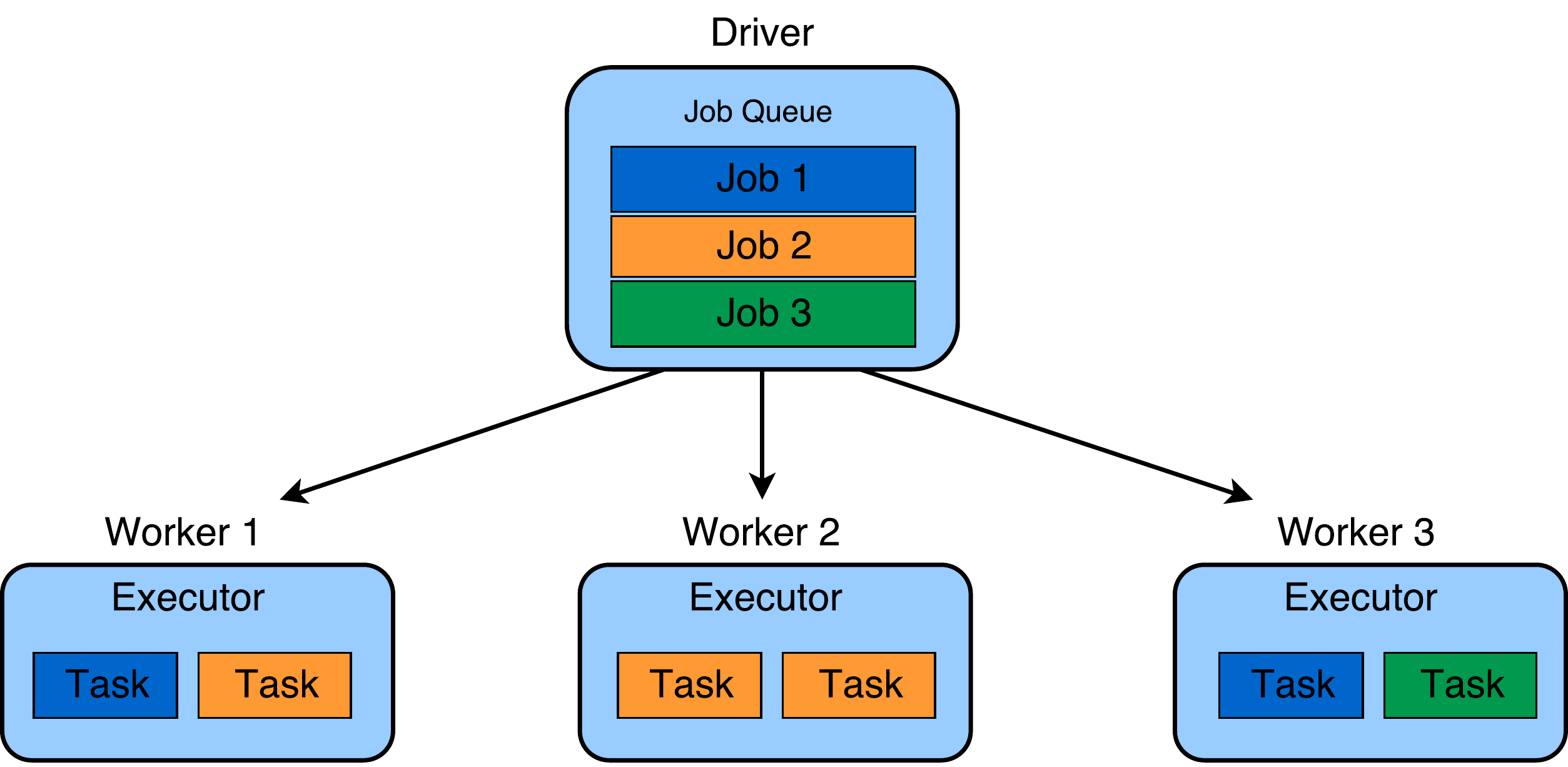}
\caption{An illustration of the dependencies of two pipelines. After the distance indexing table is constructed in parallel, Spark will broadcast it to all nodes and in the next pipeline, the executors can look up the table and fetch $E+1$ nearest neighbors quickly.}
\label{fig:SynchronousPipelines}
\end{figure}

After the pipeline is created to run CCM, a job is generated in the master node and then submitted to the cluster and partitioned into many tasks running in the executors of worker nodes. This setting is defined in the job submission and is, in general, constant. If we perform two pipelines one after other, they will always be executed sequentially. As such, we can adopt the asynchronous mechanisms to increase the parallelism and execute different pipelines concurrently. \textit{FutureAction} is the Spark API to undertake asynchronous job submission. It provides a native way for the program to express concurrent pipelines without having to deal with the detailed complexity of explicitly setting up multiple threads. In this way, we can achieve running various combinations of the parameters ($L$, $\tau$, and $E$) in parallel by executing multiple concurrent pipelines.

\section{Experiment Results}

The baseline scenario of parameters, with input time series size of 4000, $r$ of 500, $L$ with values $[500, 1000, 2000]$, $E$ and $\tau$ both with $[1,2,4]$, is set for the comparison in the experiments. In the following experiments, the Spark parallel version of CCM will be run three times on the Google Cloud Platform to obtain the average computation time. The cluster setting is 1 master node and 5 worker nodes with 4 cores CPU and 15 GB Memory.
% \begin{table}[h]\caption{Baseline setting}
% 	\centering
% 	\begin{tabular}{c|c}
% 		\hline
% 		Parameter 	      & Value           \\
% 		\hline
% 		Time series size  &    4000                     \\
% 		\hline
% 		$r$               &    500                     \\
% 		\hline
% 		$L$               & $[500, 1000, 2000]$         \\
% 		\hline
% 		$E$               &  $[1, 2, 4]$                \\
% 		\hline
% 		$\tau$            &  $[1, 2, 4]$                \\
% 		\hline
% 	\end{tabular}
% 	\label{tab:baseline}
% \end{table}

\subsection{Overview of Improvements}

This experiment compares the performance improvement of different implementations on the baseline scenario. These implementations in Table~\ref{tab:implementations} are submitted on Yarn Mode and Local Mode, separately. Yarn Mode, or cluster mode will exploit all the worker nodes existing in the cluster while Local Mode only runs applications on the master node (Single Machine). 

\begin{table}
\caption{Implementation Levels}
	\centering
	\begin{tabular}{c|c}
		\hline
		           &  Implementation Level                               \\
		\hline
		 Case A1    & Single-threaded CCM (no RDD \& Pipeline)    \\
		\hline
		 Case A2    & Synchronous CCM Transform Pipelines                              \\
		\hline
		 Case A3    &  Asynchronous CCM Transform Pipelines               \\
		\hline
		 Case A4    & Synchronous Distance Indexing Table \& CCM Transform Pipelines           \\
		\hline
		 Case A5    &  Asynchronous Distance Indexing Table \& CCM Transform Pipelines   \\
		\hline
	\end{tabular}
	\label{tab:implementations}
\end{table}

\begin{figure}[h]
\centering
\includegraphics[height=4.5cm]{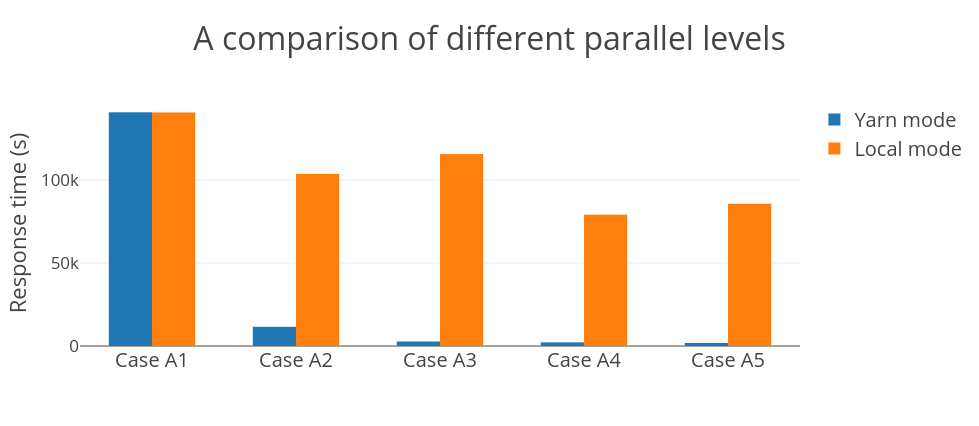}
\caption{Yarn Mode utilizes all worker nodes in the cluster, while Local Mode only run experiments on the master node. Yarn Mode significantly diminishes the average computation time of the parallel version of CCM with the help of worker nodes.}
\label{fig:parallellevels}
\end{figure}

The results are shown in Fig~\ref{fig:parallellevels}. Several conclusions can be drawn from the experimental results for different levels of parallel implementation. Firstly, the single-threaded version of CCM imposes a heavy computational cost, and there is no difference between two modes as they do not utilize the worker nodes in the cluster. Next, asynchronous pipelines can only reduce computation time in Yarn mode. After the comparison of the CPU utilization rates, it indicates that the asynchronous pipelines cannot offer more parallelization when the CPU utilization already reaches full throttle. However, when run with Yarn, the worker nodes still have rooms to improve the utilization rates. Also, as seen from the results, the spark full parallel version (\textit{Case A5}) is approximately 1.2\% the running time of the single-threaded version. Ultimately, the most significant improvement of marginal computation performance lies in adding the distance indexing table pipeline based on the CCM Transform pipeline. It reduces the computation time cost by over 80\% relative to the baseline. Such considerable improvement shows the parallel version of CCM benefits strongly from establishing the distance indexing table globally for nearest neighbors searching pipeline.

When comparing current existing public CCM implementation, rEDM R package, which created by the Hao Ye et al. \cite{ye2016redm} using lower level language C++, our Spark parallel implementation (\textit{Case A5}) is approximately 15x faster than rEDM for baseline scenario on current cluster setup on Google Compute Platform. Obviously, the parallel version can perform more favorably with a more powerful cluster (vertical scaling) or adding more workers (horizontal scaling).

\subsection{Elasticity Analysis}

\begin{table}[h]
\caption{ Elasticity Analysis}
	\centering
	\begin{tabular}{c|c|c|c|c}
		\hline
		   Parameter varied & parameter       &  Case B1     &   Case B2          &   Case B3          \\
		\hline
	    \multirow{2}{*}{$L$} & $L$ &  500       &   1000            &   2000  \\
	   \cline{2-5}
		 & others & \multicolumn{3}{c}{the same as baseline scenario} \\
		\hline
		
		\multirow{2}{*}{$E$} & $E$&  1       &   2           &   4  \\
	   \cline{2-5}
		 &others & \multicolumn{3}{c}{the same as baseline scenario} \\
		\hline
		
		\multirow{2}{*}{$\tau$} & $\tau$&  1       &   2           &   4  \\
	   \cline{2-5}
		 &others & \multicolumn{3}{c}{the same as baseline scenario} \\
		\hline
		
	\end{tabular}
	\label{tab:parameterstudies}
\end{table}

\begin{figure}[h]
\centering
\includegraphics[height=4.5cm]{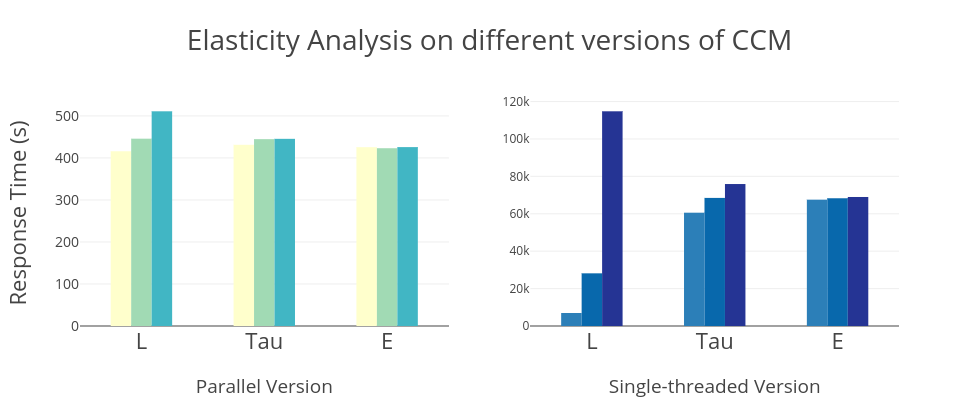}
\caption{The difference is the utility of worker nodes. The parallel version uses all of the optimization methods with five 4-core workers in the cluster, while the single-threaded version is only executed on the master node without any parallel optimization.}
\label{fig:parameterstudies}
\end{figure}

As a range of parameter settings been looped over for the best results to infer causality, testing the elasticity of running time concerning a given parameter value is necessary. Two versions of CCM (Parallel Version is the implementation \textit{Case A5} which has all degrees of parallelism, while Single-threaded Version is the implementation \textit{Case A1} which has not implemented any pipeline) will be tested with the parameter settings as shown in Table~\ref{tab:parameterstudies}. Intuitively, these cases vary only one parameter from the baseline for comparison. When doubling parameter $L$, the average run time increases to 4.06x using the Spark single-threaded version, and 1.11x using the Spark parallel version. Similarly, doubling parameter $\tau$ and $E$ almost has no impact on running time in the parallel version. However, doubling $\tau$ indeed increases the running time to 1.13x in the single-threaded version as it increase the dimension of shadow manifolds $M_x$ and $M_y$. 

In summary, the values of these parameters, especially for $L$, do influence execution time for both the single-threaded and parallel versions; however, with the current optimization of the parallel methods by introducing indexing table before nearest neighbor searching, the relative impact shrinks, which make testing relatively large parameters for the causality assessment a reality.

%Double the value,  by what factor does the run time increase.
%relative proportional

%In single machine  original version -> L influence

%In Cluster    parallel version  -> L influence

%1.
%run configuration 2 (cluster) with completed parallel version and  other variables %and keep same in baseline scenario

%2. 
% run configuration 1 (cluster) with original version and  other variables and keep %same in baseline scenario

\section{Conclusion}

% the first paragraph: summary the work
The Spark framework provides relatively convenient APIs to exploit parallelism in algorithms such as CCM. This work conducted experiments demonstrating the performance benefits of exploiting the parallelism in CCM algorithm using Spark. The scalability of Spark offers considerable benefits in accelerating the execution with the support of clusters, allowing for a significant reduction in running time when adding more worker nodes into the cluster. Of critical importance for robust application of Spark, these performance gains make this algorithm a valuable modeling tool to assess causality with confidence in an abbreviated time.  Such gains are particularly important in the context of high-velocity datasets involving human behavior and exposures, such as are commonly collected in human social and sociotechnical systems.

%Limitations
While it demonstrated potential for marked speedups, this work suffers from some pronounced limitations.  Construction of the distance indexing table trades off higher space consumption for savings in computation time; for large shadow manifolds from a large value of $L$, the indexing table can require a large quantities of system memory. However, as previous study \cite{ma2014detecting} shows, CCM can produce reliable results when input time-series length is around in the order of $n = 10^3$, for which the required memory space is well under what most current hardware can offer. 

% the final paragraph: future work
% For future work, other parallel techniques such as MPI and Graphical Processing Unit (GPU)-based parallelization offer prospects of further reducing the computation time. Recently, \cite{Yuan2016} introduced a Spark implementation that supports utilization of GPU's massively parallel processing ability to achieve both high performance and high throughput. If there are GPU/CUDA hardware in a cluster, the CCM algorithm implementation demonstrated here could be further accelerated through exploiting GPUs on the top of the Spark implementation. 

\bibliographystyle{splncs04}
\bibliography{sbp}

\begin{thebibliography}{10}
\providecommand{\url}[1]{\texttt{#1}}
\providecommand{\urlprefix}{URL }
\providecommand{\doi}[1]{https://doi.org/#1}

\bibitem{cao1997practical}
Cao, L.: Practical method for determining the minimum embedding dimension of a
  scalar time series. Physica D: Nonlinear Phenomena  \textbf{110}(1-2),
  43--50 (1997)

\bibitem{dean2008mapreduce}
Dean, J., Ghemawat, S.: Mapreduce: simplified data processing on large
  clusters. Communications of the ACM  \textbf{51}(1),  107--113 (2008)

\bibitem{heylen2011fully}
Heylen, R., Burazerovic, D., Scheunders, P.: Fully constrained least squares
  spectral unmixing by simplex projection. IEEE Transactions on Geoscience and
  Remote Sensing  \textbf{49}(11),  4112--4122 (2011)

\bibitem{kantz2004nonlinear}
Kantz, H., Schreiber, T.: Nonlinear time series analysis, vol.~7. Cambridge
  university press (2004)

\bibitem{kugiumtzis1996state}
Kugiumtzis, D.: State space reconstruction parameters in the analysis of
  chaotic time series—the role of the time window length. Physica D:
  Nonlinear Phenomena  \textbf{95}(1),  13--28 (1996)

\bibitem{luke2012systems}
Luke, D.A., Stamatakis, K.A.: Systems science methods in public health:
  dynamics, networks, and agents. Annual review of public health  \textbf{33},
  357--376 (2012)

\bibitem{luo2014causal}
Luo, C., Zheng, X., Zeng, D.: Causal inference in social media using convergent
  cross mapping. In: 2014 IEEE Joint Intelligence and Security Informatics
  Conference. pp. 260--263. IEEE (2014)

\bibitem{ma2014detecting}
Ma, H., Aihara, K., Chen, L.: Detecting causality from nonlinear dynamics with
  short-term time series. Scientific reports  \textbf{4}, ~7464 (2014)

\bibitem{maillo2017knn}
Maillo, J., Ram{\'\i}rez, S., Triguero, I., Herrera, F.: knn-is: An iterative
  spark-based design of the k-nearest neighbors classifier for big data.
  Knowledge-Based Systems  \textbf{117},  3--15 (2017)

\bibitem{monster2017causal}
M{\o}nster, D., Fusaroli, R., Tyl{\'e}n, K., Roepstorff, A., Sherson, J.F.:
  Causal inference from noisy time-series data—testing the convergent
  cross-mapping algorithm in the presence of noise and external influence.
  Future Generation Computer Systems  \textbf{73},  52--62 (2017)

\bibitem{pearl2009causal}
Pearl, J., et~al.: Causal inference in statistics: An overview. Statistics
  surveys  \textbf{3},  96--146 (2009)

\bibitem{reyes2015big}
Reyes-Ortiz, J.L., Oneto, L., Anguita, D.: Big data analytics in the cloud:
  Spark on hadoop vs mpi/openmp on beowulf. Procedia Computer Science
  \textbf{53},  121--130 (2015)

\bibitem{sugihara2012detecting}
Sugihara, G., May, R., Ye, H., Hsieh, C.h., Deyle, E., Fogarty, M., Munch, S.:
  Detecting causality in complex ecosystems. science  \textbf{338}(6106),
  496--500 (2012)

\bibitem{takens1981detecting}
Takens, F.: Detecting strange attractors in turbulence. In: Dynamical systems
  and turbulence, Warwick 1980, pp. 366--381. Springer (1981)

\bibitem{vavilapalli2013apache}
Vavilapalli, V.K., Murthy, A.C., Douglas, C., Agarwal, S., Konar, M., Evans,
  R., Graves, T., Lowe, J., Shah, H., Seth, S., et~al.: Apache hadoop yarn: Yet
  another resource negotiator. In: Proceedings of the 4th annual Symposium on
  Cloud Computing. p.~5. ACM (2013)

\bibitem{verma2016analysis}
Verma, A.K., Garg, A., Blaber, A., Fazel-Rezai, R., Tavakolian, K.: Analysis of
  causal cardio-postural interaction under orthostatic stress using convergent
  cross mapping. In: 2016 38th Annual International Conference of the IEEE
  Engineering in Medicine and Biology Society (EMBC). pp. 2319--2322. IEEE
  (2016)

\bibitem{ye2016redm}
Ye, H., Clark, A., Deyle, E., Sugihara, G.: redm: an r package for empirical
  dynamic modeling and convergent cross-mapping  (2016)

\bibitem{zaharia2012resilient}
Zaharia, M., Chowdhury, M., Das, T., Dave, A., Ma, J., McCauley, M., Franklin,
  M.J., Shenker, S., Stoica, I.: Resilient distributed datasets: A
  fault-tolerant abstraction for in-memory cluster computing. In: Proceedings
  of the 9th USENIX conference on Networked Systems Design and Implementation.
  pp.~2--2. USENIX Association (2012)

\bibitem{zaharia2010spark}
Zaharia, M., Chowdhury, M., Franklin, M.J., Shenker, S., Stoica, I.: Spark:
  Cluster computing with working sets. HotCloud  \textbf{10}(10-10), ~95 (2010)

\end{thebibliography}

\end{document}